\documentclass[journal]{IEEEtran}

\usepackage{amsmath,amssymb,graphicx,cite}
\usepackage{color}
\usepackage{enumitem}
\usepackage{hyperref}
\usepackage{multirow}
\usepackage{siunitx}
\usepackage{booktabs}
\usepackage{balance}
\usepackage{bm}

\def\db{\textrm{\normalfont dB}}
\DeclareDocumentCommand \Lframe{ m g }{{\mathcal{L}_{\operatorname{f-#1} \IfNoValueF {#2} {,{#2}}}}}
\DeclareDocumentCommand \Lclip{ m g }{{\mathcal{L}_{\operatorname{c-#1} \IfNoValueF {#2} {,{#2}}}}}

\begin{document}
\title{Finding Strength in Weakness:\\Learning to Separate Sounds with Weak Supervision}

\author{Fatemeh Pishdadian,~\IEEEmembership{Student Member,~IEEE,}
        Gordon~Wichern,~\IEEEmembership{Member,~IEEE,}
        and~Jonathan~Le~Roux,~\IEEEmembership{Senior~Member,~IEEE}%
\thanks{F. Pishdadian is with Interactive Audio Lab, Northwestern University, Evanston, IL, USA (e-mail: fpishdadian@u.northwestern.edu). G. Wichern and J. Le Roux are with Mitsubishi Electric Research Laboratories (MERL), Cambridge, MA, USA (e-mail: \{wichern,leroux\}@merl.com).}%
\thanks{This work was performed while F.~Pishdadian was an intern at MERL.}}

\markboth{IEEE/ACM Transactions on Audio, Speech, and Language Processing}%
{Pishdadian \MakeLowercase{\textit{et al.}}: Finding Strength in Weakness}

\IEEEpubid{This article appears in IEEE/ACM Transactions on Audio, Speech, and Language Processing, Volume 28, pages 2386 - 2399 \textcopyright 2020 IEEE.}

\maketitle

\begin{abstract}
While there has been much recent progress using deep learning techniques to separate speech and music audio signals, these systems typically require large collections of isolated sources during the training process.  When extending audio source separation algorithms to more general domains such as environmental monitoring, it may not be possible to obtain isolated signals for training.  Here, we propose objective functions and network architectures that enable training a source separation system with weak labels. In this scenario, weak labels are defined in contrast with strong time-frequency (TF) labels such as those obtained from isolated sources, and refer either to frame-level weak labels where one only has access to the time periods when different sources are active in an audio mixture, or to clip-level weak labels that only indicate the presence or absence of sounds in an entire audio clip.  
We train a separator that estimates a TF mask for each type of sound event, using a sound event classifier as an assessor of the separator's performance to bridge the gap between the TF-level separation and the ground truth weak labels only available at the frame or clip level. Our objective function requires the separator to estimate a source such that the classifier applied to it will assign high probability to the class corresponding to that source and low probability to all other classes.
The objective function also enforces that the separated sources sum up to the mixture. We benchmark the performance of our algorithm using synthetic mixtures of overlapping events created from a database of sounds recorded in urban environments, and show that the method can also be applied to other tasks such as music source separation.  Compared to training a network using isolated sources, our model achieves somewhat lower but still significant SI-SDR improvement, even in scenarios with significant sound event overlap.
\end{abstract}


\IEEEpeerreviewmaketitle

\section{Introduction}
\label{sec:intro}

\IEEEPARstart{A}{udio} 
source separation aims to isolate individual sound sources in a complex auditory scene. This process plays an essential role in a variety of applications, including speech recognition in noisy environments \cite{ehlers1997blind}, speaker identification in a multi-speaker scenario \cite{cooke2010monaural}, and music remixing \cite{woodruff2006remixing}. 

Time-frequency (TF) domain mask inference is a common approach to solving the under-determined source separation problem, in which the number of audio sources exceeds the number of recorded channels~\cite{lyon1983computational, weintraub1985theory, wang2006CASA}. In such an approach, a raw audio mixture is first transformed into an intermediary representation, e.g., the short-time Fourier transform (STFT). Each source is then estimated by applying a weighting function with values typically in $[0,1]$, referred to as a \emph{mask}, to the mixture in the transform domain before converting back to the time domain.

\IEEEpubidadjcol

Supervised mask inference methods, especially those using deep neural networks, have gained much popularity over the past decade, due to their successful performance in speech enhancement~\cite{weninger2014single,Weninger2014GlobalSIP12, Erdogan2015, wang2018supervised}, speech separation~\cite{hershey2016deep,Isik2016, Kolbaek2017,Wang2018ICASSP04Alternative,luo2019convTasNet}, music separation~\cite{luo2017deep, Takahashi2018, seetharaman2019class, manilow2019slakh, stoter19,kumar2018music}, and sound effect separation~\cite{kavalerov2019universal}. 
These approaches typically require a large dataset of isolated sound sources to construct training targets for estimating the TF masks from the corresponding sound mixtures.  However, obtaining the isolated sources that compose a mixture may be expensive, require complicated recording setups, or necessitate the creation of synthetic mixtures that lack a certain amount of realism. In the extreme case, some sounds may never be recorded in isolation, such as the sound of a specific machine part that only occurs when a machine is running and other parts might also be making some sound. 

In cases where isolated sources are not available for training the separation system, it is also unrealistic for humans to use signal processing tools to manually label the audio at the granularity level of TF bins, especially to do so accurately and at scale.  However, it is reasonable to assume that they can produce limited labels for the activity of sound sources within some time range.  Even non-expert users have successfully provided labels for musical instrument detection~\cite{humphrey2018openmic} and sound event detection (SED)~\cite{cartwright2019crowdsourcing}, where the labels consisted of the type of audio events as well as the precise time of their occurrence in a given recording. The annotation burden can be further reduced, as has been considered in the SED task, by replacing the fine resolution labels on the precise sound event onsets and offsets (typically defined at a resolution of a few dozens milliseconds) by a coarse temporal label indicating the presence or absence of a sound event within a particular audio clip (e.g., on the order of 10 s). Since the fine resolution labels are typically defined at the level of an STFT frame, we hereafter refer to them as frame-level labels, while we refer to the coarse labels as clip-level labels.

In this paper, our goal is to investigate whether separation methods using deep learning, which are typically trained in a fully supervised setup using TF-bin-level labels, can be trained using weaker (frame-level or clip-level) labels. 
We thus attempt to perform, similarly to what has been done in the context of SED, a transition from strong to weak labels.
We shall however point out an important difference regarding the notion of strength of a label in the context of SED and separation.
In SED, the goal is to estimate the type of an audio event together with its precise onset and offset, with the corresponding ground truth referred to as a strong label. In contrast, ground truth limited to presence or absence of a sound within a coarser time range is referred to as a weak label. 
We refer the interested reader to the description of a weakly labeled SED task in the DCASE 2017 challenge~\cite{mesaros2019sound}. 
In the context of source separation, complete ground truth consists in each source's isolated signal, which amounts to having information on each source at the granularity of a TF bin. Strong labels for SED are thus only weak labels for source separation. To our knowledge, no deep-learning-based source separation system has so far been presented that can be trained under the assumption of sole availability of frame-level labels (let alone clip-level ones) and is able to separate mixtures at test time without side information.

Weakly labeled SED approaches typically leverage multiple instance learning, where an instance-level (i.e., fine temporal resolution) predictor is trained by aggregating or pooling the instance-level predictions to match the labels at the ``bag'' level (i.e., a chunk of audio on the order of several seconds, and its associated coarse ground truth label). We would like to use a similar concept for source separation, where the instance-level prediction is now at the level of the TF bin, and the bag level is either that of a frame or that of a whole clip.
A different approach to pooling is however needed in
SED and source separation systems.  
In weakly-labeled SED, consecutive time frames will often share the same class labels. In weakly-labeled separation, on the other hand, the structure is much more intricate,
as frequency bins sharing the same label may be far from each other, often harmonically spaced in a highly variable manner even among the same types of sounds.

To overcome these difficulties in pooling over the frequency and time dimensions, we propose a form of discriminative pooling, where an SED classifier is employed as the principal metric for loss calculation when training the separator.  When applied to a separated source, the classifier is expected to detect that only a single class is present, while all other sources are inactive. 
Furthermore, we propose a multi-task learning approach in training the separator, combining the audio event classification objective with an additional separation-specific objective that enforces the separated sources to sum up to the mixture.  Our model learns to separate based solely on weak labels, either at the frame level or at the clip level. Clip-level labels are equivalents of SED weak labels, which do not require the sound to be active throughout the entire time period for which the label applies. In this work, we extend~\cite{pishdadian2020learning} to further investigate the contribution of the classification and separation objective function terms to the quality of learned masks, as well as the correlation between classifier and separator performance. We also explore different training strategies, where the classifier and separator are trained jointly, or we first train the classifier on the mixtures, and then fix or fine-tune its weights while training the separator.  Empirical comparison of weakly-labeled separation performance to the strongly-labeled case (when isolated sources are available) is carried out using synthetic mixtures created from the \emph{UrbanSound8K} \cite{SalamonUrbanSoundACMMM14} and \emph{MUSDB18}~\cite{musdb18} datasets.

\textbf{Related work:} As previously mentioned, there has been a resurgence in multiple instance learning approaches for audio following the DCASE 2017 challenge~\cite{mesaros2019sound}, where several studies examine deep network architectures~\cite{kumar2016audio, su2017weakly, xu2018large} and/or pooling functions~\cite{mcfee2018adaptive, wang2019comparison, kimsound} for the weakly-labeled SED task.  There have also been several applications of multiple instance learning for music, including detecting instruments in mixtures~\cite{little2008learning}, applying artist-/album-level labels to individual tracks~\cite{mandel2008multiple}, and saliency-based singing voice detection~\cite{schluter2016learning}.  

Deep learning based techniques are currently dominant for fully supervised source separation, and typically trained to separate a single source class of interest, such as vocals or a particular instrument type from music mixtures~\cite{uhlich2017improving, Takahashi2018, manilow2019slakh}, or speech from noise~\cite{Erdogan2015, wang2019comparison}.  An alternative class of techniques such as deep clustering~\cite{hershey2016deep} and permutation invariant training~\cite{hershey2016deep,Kolbaek2017} is required when the source types to be separated are very similar, e.g., separating speech from speech.  
The fully supervised approaches most relevant to the current study are those that train a single network to separate multiple classes of musical instruments~\cite{kumar2018music, seetharaman2019class, slizovskaia2019end}.

Semi-supervised separation methods based on generative adversarial learning were proposed in \cite{zhang2017weakly, stoller2018adversarial}. The key assumption of these methods is that estimated sources produced by an optimal separator should be indistinguishable from real sound sources, i.e., they should be samples drawn from the same distribution. 
These adversarial approaches are semi-supervised, since one-to-one correspondence between the mixtures and the real isolated sources used for training is not required. Nevertheless, their training is indeed dependent on the existence of some dataset of isolated sources. 
However, the need for isolated data can be relaxed when separating a single type of source while only observing isolated background and the target source in the background~\cite{stowell2015denoising, michelashvili2019semi}.  Another class of source separation techniques based on weak labels assumes the availability of weak labels at both training and inference time, such as the score-informed approach in~\cite{ewert2017structured}, the variational auto-encoder in~\cite{karamatli2019audio}, and the audio-visual approach in~\cite{gao2019co}, where the video provides (weak) class  labels to guide the audio separation. Our approach can separate multiple source classes, does not require seeing any sources in isolation, and requires only the audio mixture (no labels) during inference.

Another line of research performs source separation implicitly when training SED systems using either NMF~\cite{heittola2011sound} or deep networks~\cite{kong2019sound}.  The method in~\cite{kong2019sound} is composed of two stages: first, a segmentation mapping is applied to the TF representation of an audio recording to obtain TF segmentation masks, then a classification mapping is applied to the segmentation masks to estimate the presence probability of sound events. 
The authors suggest that the separation task can be performed as a byproduct of event detection using the learned segmentation masks. However, their objective function is only event detection cross-entropy and does not include any terms modeling the separation problem explicitly, such as enforcing each separated mask to belong only to a single source, or enforcing estimated sources to sum up to the mixture as in our approach.  Furthermore, they test their method only on isolated sources in background noise, whereas our experiments deal with multiple overlapping sound events. 

\section{Joint Separation-Classification Approach}
\label{sec:sys-arch}

We take a joint separation-classification approach to audio source separation through weak labels. In this section, we first
provide basic definitions for the mask-based single-channel source separation problem and briefly review the fully supervised separation setup. We then present our weakly supervised separation model, formulate the objective function, and discuss the training setup in detail.

\subsection{Mask-based Single-channel Audio Source Separation}

Throughout this work, we assume a monaural source separation scenario, where only one recording channel of the mixture is available. 
We observe a mixture
\begin{equation}\label{mix-model}
    x(t) = \sum_{i=1}^{n} s_i(t),
\end{equation}
where $x(t)$ and $s_i(t)$ respectively denote the mixture signal and the $i$-th sound source signal in the time domain, and $n$ is the total number of sound sources in the mixture. 
Note that each \emph{sound source} is here assumed to belong to a distinct \emph{sound class} (e.g., musical instrument, human speech, dog bark, etc.), in other words all instances of the same sound class are considered as a single sound source. We thus use these two terms interchangeably hereafter.

As mentioned earlier, a common approach to solving the under-determined separation problem is to perform masking on the mixture in some time-frequency (TF) domain,
where there is less overlap between sources than in the time domain. 
We denote the magnitude TF representation (e.g., magnitude STFT) of the mixture by $X_{\omega, \tau}$, where $\omega$ and $\tau$ are frequency and time-frame indices, respectively. 

The first step in a typical TF-masking-based method is to estimate
source magnitudes by performing element-wise multiplication of the mixture magnitude with a set of estimated masking functions.
Let $\hat{M}_{i, \omega, \tau}$ denote a TF mask estimate for the $i$-th source, taking on values in $[0,1]$, with $\hat{M}_{i, \omega, \tau}$ being ideally very close or equal to $0$ where the source is inactive and very close or equal to $1$ where the source is dominant in the mixture.
The masking operation can be formulated as
\begin{equation}\label{est-src}
    \hat{S}_{i, \omega, \tau} = \hat{M}_{i, \omega, \tau} X_{\omega, \tau},
\end{equation}
where $\hat{S}_{i, \omega, \tau}$ is the estimated magnitude of the $i$-th source.
The estimated source magnitudes are then typically combined with the mixture phase and converted back to the time domain through an inverse transform (e.g., iSTFT). We leave extensions of our method that consider estimation of the phase or the complex spectrogram of the sources to future work.

\subsection{Fully Supervised Separation}
\label{sec:sup_mi}
The supervised mask inference task aims at training a model to generate estimates of the sources present in a given audio mixture via the estimation of masks to be applied to a TF representation of the mixture.

In the fully supervised separation scenario, the time-domain signals of the isolated sources, their  TF-domain representations, or TF masks built from them (e.g., the ideal binary mask or the ideal ratio mask \cite{Erdogan2015}) are used as targets in model training. 
We refer to such targets as ``strong labels,'' as they provide information about sound classes at the TF-bin level. 

Various loss functions have been used in fully supervised mask inference, such as mask approximation (MA), magnitude spectrum approximation (MSA), phase spectrum approximation (PSA), and waveform approximation (WA) \cite{Erdogan2015,leroux2019phasebook}. We here focus on the MSA objective with $L^1$ norm for simplicity:
\begin{eqnarray}
    \mathcal{L}_{\mathrm{mi}} &=& \sum_{i,\omega,\tau} | \hat{S}_{i,\omega, \tau} - S_{i,\omega, \tau} | \nonumber\\
    &=& \sum_{i,\omega,\tau} |X_{\omega, \tau} \hat{M}_{i,\omega, \tau} - S_{i,\omega, \tau} |, \label{fully_supervised_mi}
\end{eqnarray}
where $S_{i,\omega, \tau}$ denotes the magnitude TF representation of the $i$-th isolated source and $|\cdot|$ indicates the modulus operator.

In the weakly supervised scenario, we no longer have access to TF-bin-level labels (target sources or masks). The target labels instead indicate only the sound class presence at the frame or clip level.
The next two sections present our approach to training mask inference networks using only frame- or clip-level sound labels.

\begin{figure*}[t]
  \centering
 \includegraphics[width=0.85\textwidth]{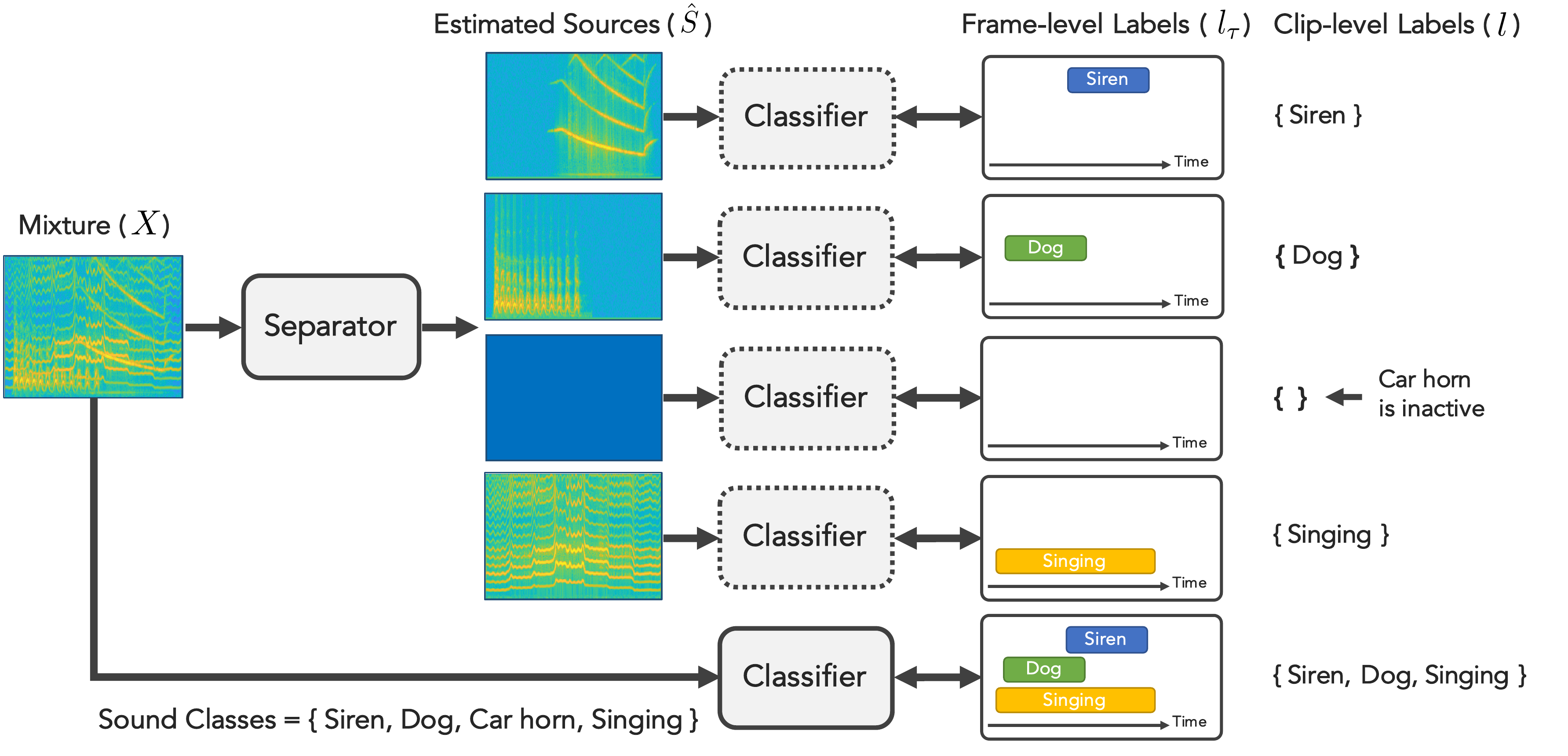}
\caption{The joint separation-classification model. The separator receives an audio mixture and returns source estimates (the blue square is the estimate of an inactive source).  The classifier processes separately the mixture and each estimated source (dashed lines indicate shared parameters). When applied to the mixture, the classifier should output the presence probabilities for all classes. The separator is trained such that if any of the source estimates is used as input to the classifier, the classifier output is the presence probability for that source along with zeros for all other sources. }
\label{fig:model_diagram}
\end{figure*}

\subsection{Weakly Supervised Separation}
\label{sec:jsc-model}

At a high level, our model is composed of two main blocks: a source separator and an audio event classifier. The block diagram of the entire system is shown in Fig.~\ref{fig:model_diagram}. 

The separator block receives the TF representation (e.g., magnitude STFT) of a mixture and outputs estimates $\hat{\bm{S}}_i$, $i=1,\dots,n$, for each of the sources in the TF domain,
where $n$ indicates the total number of sound classes available in a dataset. 
We assume the number of active sound classes in a given mixture ranges from 1 to $n$. 

The input to the classifier block is also a TF representation. 
In general, the TF representation used as input to the classifier may be of a different type from the separator output (and the separator input, as we typically assume they are in the same domain), as long as we can pass gradients through the transform used to compute it. For instance, the classifier input can be a mel spectrogram while the separator input and output are a magnitude STFT.
Given a TF representation $\bm{Y}$ as input, the classifier computes frame-level class probabilities $p_{i,\tau}(\bm{Y})$ for each source class $i$ and time frame $\tau$, representing how likely each source class is to be active at each time frame in $\bm{Y}$, or clip-level class probabilities $p_{i}(\bm{Y})$ for each source class $i$, representing how likely each source class is to be active (anywhere) within $\bm{Y}$. 

We denote the frame-level label for the $i$-th sound source at frame $\tau$ by $l_{i,\tau}$, which indicates whether the source is active at that frame ($l_{i,\tau}=1$) or not ($l_{i,\tau}=0$). The clip-level label $l_i$ indicates whether the source $i$ is present anywhere in the clip, i.e., $l_i=1$ would correspond to the case where $l_{i,\tau}=1$ for at least one time-frame $\tau$, and $l_i=0$, otherwise, assuming that frame-level labels were available.
Note that $l_{i,\tau}$ may be regarded as the output of a pooling operation across frequency applied to the TF-level labels for the $i$-th isolated source at frame $\tau$, while $l_{i}$ would be further pooled across time. 

Our main idea is that, if we can train a classifier that performs well in predicting source class activities on natural mixtures, where sound classes may sometimes occur in isolation and other times overlap with other classes, we can use that classifier as a critic of the separator's performance, assessing how well the separator is able to separate each source. We can thus use weak labels, either at the frame or clip level, to train the separator through the classifier.
For instance, if source $i$ is active at frame $\tau$, passing the estimated source $\hat{\bm{S}}_i$ as input to the classifier should result in classification outputs such that $p_{i,\tau}(\hat{\bm{S}}_{i}) = 1$ and $p_{j,\tau}(\hat{\bm{S}}_i) = 0$ for all $j \neq i$, because all other sources should be removed from $\hat{\bm{S}}_i$. 
This is illustrated in Fig.~\ref{fig:model_diagram}, where we have shown both frame-level labels, with onsets and offsets for each sound class, and clip-level labels where only presence or absence within a clip is indicated.

The classifier can be trained independently or jointly with the separator. The separator, on the other hand, requires the classification results while training, since TF-bin labels are not available and the classifier is required to pool over the TF-bin-level source activity predictions to make predictions at each time frame or for the whole clip.  In this work, we consider three training strategies: i) training the separator and classifier jointly from scratch, ii) training the separator through a pre-trained classifier while the classifier is being fine-tuned, and iii) training the separator through a pre-trained and fixed classifier. It should be noted that we pre-train the classifier only on mixtures, not on isolated sources, as the latter case would violate the assumption that strong labels are unavailable. 

\subsection{Weakly Supervised Objective Function}
\label{sec:objective}
\subsubsection{Mixture loss}
Our principal goal in training the model is to achieve high quality separation, which requires explicit optimization of mask estimates, even if ground truth TF labels are not available. 
To this end, a key constraint is to enforce the
output signals of the separator to add up to explain the input mixture.
Indeed, this constraint is critical in preventing the separator from producing masks that solely focus on the most discriminating time-frequency components for classification without fully reconstructing the entire source. We can promote the enforcement of this constraint through a mixture loss term in the objective function that minimizes the discrepancy between the mixture and the sum of estimated source spectrograms, or between the mixture magnitude and the sum of estimated source magnitudes, assuming that all source estimates are obtained using the mixture phase.
A vanilla version of such a term can be formulated using an $L^1$ loss as
\begin{equation}\label{sep-loss}
\mathcal{L}_{\mathrm{mix},\mathrm{vanilla}} =  \sum_{\omega,\tau} |X_{\omega, \tau} - \sum_{i=1}^n \hat{S}_{i,\omega, \tau}|.
\end{equation}
Thanks to the information provided by the weak labels, we can in fact further enforce that only the sum over active sources should be equal to the mixture, and all inactive sources should be silent. In the frame-level case, the vanilla loss term in \eqref{sep-loss} can therefore be replaced by a more explicitly constrained version defined in two parts:
\begin{align}\label{sep-loss-frame}
  \mathcal{L}_{\operatorname{f-mix}} = \sum_{\omega,\tau}|X_{\omega,\tau} - \sum_{i \in \mathcal{A_{\tau}}} \hat{S}_{i,\omega,\tau} | 
  +  \sum_{\omega,\tau} \sum_{i \notin \mathcal{A_{\tau}} } | \hat{S}_{i,\omega,\tau}|,
\end{align}
where $\mathcal{A_{\tau}}$ is the set of active source indices in time frame $\tau$.  Moreover, given the weak labels, we can locate mixture frames where no sources are active and exclude those entirely from loss computation. We refer to $\mathcal{L}_{\operatorname{f-mix}}$ as the frame-level mixture loss.
In the clip-level case, we only have information about the set $\mathcal{A}$ of active source indices for the whole clip, which we can use to similarly modify \eqref{sep-loss} as:
\begin{align}\label{sep-loss-clip}
  \mathcal{L}_{\operatorname{c-mix}} = \sum_{\omega,\tau}|X_{\omega,\tau} - \sum_{i \in \mathcal{A}} \hat{S}_{i,\omega,\tau} | 
  +  \sum_{\omega,\tau} \sum_{i \notin \mathcal{A} } | \hat{S}_{i,\omega,\tau}|,
\end{align}
which we refer to as the clip-level mixture loss.
In our experiments, these refinements to the vanilla mixture loss in \eqref{sep-loss} proved very important for obtaining good mask estimates.

\subsubsection{Frame-level loss}
The sound classes identified by the classifier should match the correct labels,
whether it is applied to the input mixture or any of the sources estimated by the separator. This can be achieved by including a binary cross-entropy term between the classifier output and the corresponding true labels.
Let $H(l,p)$ denote the binary cross-entropy function defined as 
\begin{equation}
    H(l, p) = - l \log(p) - (1-l) \log(1-p),
\end{equation}
where $l \in [0,1]$ and $p \in [0,1]$ respectively denote the true and estimated class probabilities.
We denote by $\Lframe{class}(\bm{Y}, \tau)$ the frame-level classification loss at frame $\tau$ for an input spectrogram $\bm{Y}$ and its associated frame-level weak labels (where labels are left implicit for simplicity of notation). This loss is computed on the mixture $\bm{X}$ and on each separated source $\hat{\bm{S}}_i$.
For the mixture $\bm{X}$, the classification loss at frame $\tau$ can be classically computed as the sum of binary cross-entropy terms over all sources,
\begin{equation}\label{frame_class_mix}
    \Lframe{class}(\bm{X}, \tau) = \sum_{i=1}^n H(l_{i,\tau}, p_{i,\tau}(\bm{X})),
\end{equation}
where $l_{i,\tau} \in \{0,1\}$ is the true frame-level label for the $i$-th source at frame $\tau$.
For the $i$-th estimated source $\hat{\bm{S}}_i$, the associated labels at each frame $\tau$ are obtained from the labels for mixture $\bm{X}$ by keeping only the label $l_{i,\tau}$ for the $i$-th source, whose activity should be the same as in $\bm{X}$, and replacing the labels for all other sources with zeros, as they should now be inactive.
The loss is thus computed as:
\begin{equation}\label{frame_class_est}
    \Lframe{class}(\hat{\bm{S}}_i,\tau) =  H(l_{i,\tau}, p_{i,\tau}(\hat{\bm{S}}_i)) +  \sum_{j \neq i} H(0, p_{j,\tau}(\hat{\bm{S}}_i)).
\end{equation}

The total frame-level classification loss $\mathcal{L}_{\operatorname{f-class}}^{\operatorname{total}}(\tau)$ at frame $\tau$, 
where the classifier is applied to the mixture and all the estimated sources, is computed as
\begin{equation}\label{class-total}
    \mathcal{L}_{\operatorname{f-class}}^{\operatorname{total}}(\tau) = 
    \Lframe{class}(\bm{X},\tau) + \sum_{i=1}^n \Lframe{class}(\hat{\bm{S}}_i,\tau).
\end{equation}

Combining the mixture loss and the classification loss, the overall frame-level loss function to be minimized can be written as
\begin{equation}\label{frame_loss}
    \mathcal{L}_{\operatorname{f}} =
    \sum_{\tau} 
    \mathcal{L}_{\operatorname{f-class}}^{\operatorname{total}}(\tau) + \alpha \mathcal{L}_{\mathrm{f-mix}},
\end{equation}
where $\alpha \geq 0$ is a tunable parameter controlling the contribution of the mixture loss to the total loss. 

\subsubsection{Clip-level loss}
When only clip-level weak labels are available, we assume that the classifier outputs a single prediction at the clip level. For example, in our experiments, a time-pooling operation is applied to the output of the frame classifier to map frame labels to clip labels as is commonly done in the weakly-labeled SED literature~\cite{mcfee2018adaptive, wang2019comparison} (see Section \ref{sec:net-arch}).
The classification loss given the clip-level labels is formulated as
\begin{align}\label{class-total-clip}
    \mathcal{L}_{\operatorname{c-class}}^{\operatorname{total}} &= 
    \Lclip{class}(\bm{X}) + \sum_{i=1}^n \Lclip{class}(\hat{\bm{S}}_i),
\end{align}
with
\begin{align}
    \Lclip{class}(\bm{X}) &=
    \sum_{i=1}^n H(l_{i}, p_{i}(\bm{X})), \\
    \Lclip{class}(\hat{\bm{S}}_i)&= H(l_{i}, p_{i}(\hat{\bm{S}}_i)) + \sum_{j \neq i} H(0, p_{j}(\hat{\bm{S}}_i)),
\end{align}
where $l_i$ denotes the clip-level label for the $i$-th sound class and $p_i$ is the clip-level class probability output by the classifier for the $i$-th class.
Finally, the total loss in the clip-level case is computed as
\begin{equation}\label{clip_loss}
    \mathcal{L}_{\operatorname{c}} =
    \mathcal{L}_{\operatorname{c-class}}^{\operatorname{total}} + \alpha  \mathcal{L}_{\operatorname{c-mix}}.
\end{equation}

Table \ref{loss-table} summarizes the loss functions used with strong labels, frame-level weak labels, and clip-level weak labels.  In the present work, we only consider systems trained exclusively with one of the three label types, i.e., we leave the study of combining available training data with different label-types as a topic of future work. We also note that, while the different loss functions in Table~\ref{loss-table} are interrelated, the decisions they make in separating and classifying sources are not necessarily consistent with each other.  For example, a clip-level classifier using max-pooling to combine frame-level decisions could make the right decision at the clip level even though all frame-level decisions are incorrect (i.e., the correct maximum score occurs at a frame that does not have the correct frame-level label). 
\renewcommand{\arraystretch}{1.5}
\begin{table*}[htbp]
\caption{Summary of fully-supervised and weakly-supervised loss functions. }\vspace{-.3cm}
\label{loss-table}
\begin{center}
\resizebox{0.7\textwidth}{!}{
\begin{tabular}{ccl}
\toprule
Label type & Eqn. & \multicolumn{1}{l}{Loss function} \\
\cmidrule(lr){1-3}
Strong & (\ref{fully_supervised_mi}) &
    $\mathcal{L}_{\mathrm{mi}} = \sum_{i,\omega,\tau} |X_{\omega, \tau} \hat{M}_{i,\omega, \tau} - S_{i,\omega, \tau}|$\\
\cmidrule(lr){1-3}    
 \multirow{3}{*}{Frame} & (\ref{sep-loss-frame}) & $\mathcal{L}_{\operatorname{f-mix}} = \sum_{\omega,\tau}|X_{\omega,\tau} - \sum_{i \in \mathcal{A_{\tau}}} \hat{S}_{i,\omega,\tau} | 
  +  \sum_{\omega,\tau} \sum_{i \notin \mathcal{A_{\tau}} } | \hat{S}_{i,\omega,\tau}|$   \\
  &  (\ref{class-total}) & $\mathcal{L}_{\operatorname{f-class}}^{\operatorname{total}}(\tau) = \sum_{i} H(l_{i,\tau}, p_{i,\tau}(\bm{X})) + \sum_{i} ( H(l_{i,\tau}, p_{i,\tau}(\hat{\bm{S}}_i))
   +  \sum_{j \neq i} H(0,  p_{j,\tau}(\hat{\bm{S}}_i))) $ \\
  &  (\ref{frame_loss}) & $\mathcal{L}_{\operatorname{f}} =
    \sum_{\tau} 
    \mathcal{L}_{\operatorname{f-class}}^{\operatorname{total}}(\tau) + \alpha \mathcal{L}_{\operatorname{f-mix}}$\\
  \cmidrule(lr){1-3}
 \multirow{3}{*}{Clip}  & (\ref{sep-loss-clip}) & $\mathcal{L}_{\operatorname{c-mix}} = \sum_{\omega,\tau}|X_{\omega,\tau} - \sum_{i \in \mathcal{A}} \hat{S}_{i,\omega,\tau} | 
  +  \sum_{\omega,\tau} \sum_{i \notin \mathcal{A}} | \hat{S}_{i,\omega,\tau}|$\\
  &   (\ref{class-total-clip}) & $\mathcal{L}_{\operatorname{c-class}}^{\operatorname{total}} = \sum_{i} H(l_{i}, p_{i}(\bm{X})) + 
    \sum_{i} (H(l_{i}, p_{i}(\hat{\bm{S}}_i)) + \sum_{j \neq i} H(0, p_{j}(\hat{\bm{S}}_i))) $\\
 &  (\ref{clip_loss}) & $\mathcal{L}_{\operatorname{c}} =
    \mathcal{L}_{\operatorname{c-class}}^{\operatorname{total}} + \alpha  \mathcal{L}_{\operatorname{c-mix}}$\\
\bottomrule
\end{tabular}
}\vspace{-0.5cm}
\end{center}
\end{table*}

\subsection{Balancing Class Weights}\label{sec:class_weights}

In the preceding discussions, all sound sources contribute equally to the total loss value. This is a reasonable setup in cases where all sound sources are equally likely to be active at any given time. 
However, sound sources may in general occur with very different activity levels in a dataset. For instance, a dataset of urban sounds might include rare, impulsive sound events such as gun shots, as well as sounds that are active over long periods of time such as street music.  Therefore, we weight each sound class during training to balance the contribution to the total loss of active and inactive frames for that class, which also equalizes the weight between classes.

Let $\gamma_i$ denote the probability for the $i$-th source to be active at any given frame. We compute $\gamma_i$ from the training data as the ratio of the 
number of frames in the dataset where the $i$-th source is active to the 
total number of frames in the dataset. We aim at increasing the contribution of sources occurring less frequently or for very short periods of time (e.g., $\gamma_i=0.1$) in the total loss, while decreasing the contribution of sources that are active most of the time (e.g., $\gamma_i=0.9$). 
This can be achieved by weighting the loss terms corresponding to frames where a source is active by the inverse of the source's prior probability of being active, and similarly for the frames where the source is inactive. 
We define the loss weight for the $i$-th source as
\begin{equation}\label{loss-weight}
W_{i,\tau} = \left\{ 
\begin{array}{lr}
\gamma_i^{-1} & i \in \mathcal{A}_{\tau},  \\
(1-\gamma_i)^{-1}& i \notin \mathcal{A}_{\tau},
\end{array}\right.   
\end{equation}
where $\mathcal{A_{\tau}}$ is the set of active source indices in time frame $\tau$. When using such weights in a loss term, the expected number of frames contributing to that loss is not only the same for active and inactive regions of a given source, but also the same across all sources. 

We can incorporate these weights in the fully supervised mask inference loss~\eqref{fully_supervised_mi} as
\begin{equation}\label{loss-strong-src-weighted}
\mathcal{L}_{\mathrm{mi},W} 
    = \sum_{i,\omega,\tau} W_{i,\tau} \left|X_{\omega, \tau} \hat{M}_{i,\omega, \tau} - S_{i,\omega, \tau} \right|.
\end{equation}

We can also incorporate these weights in the case of frame-level weak labels, reformulating the classification loss functions from~\eqref{frame_class_mix} and~\eqref{frame_class_est} as follows:
\begin{align}
    \Lframe{class}{W}(\bm{X}, \tau) &= \sum_{i=1}^n W_{i,\tau} H(l_{i,\tau}, p_{i,\tau}(\bm{X})),\\
    \Lframe{class}{W}(\hat{\bm{S}}_i,\tau) &=  W_{i,\tau} H(l_{i,\tau}, p_{i,\tau}(\hat{\bm{S}}_i)) \nonumber\\ 
    &\hspace{0.5cm}+  \sum_{j \neq i} W_{j,\tau} H(0,  p_{j,\tau}(\hat{\bm{S}}_i)).
\end{align}
The total classification loss $\mathcal{L}_{\operatorname{f-class}}^{\operatorname{total}}(\tau)$ in \eqref{class-total} and the overall loss function $\mathcal{L}_{\operatorname{f}}$ in \eqref{frame_loss} are modified accordingly, leading to $\mathcal{L}_{\operatorname{f-class},W}^{\operatorname{total}}(\tau)$ and $\mathcal{L}_{\operatorname{f},W}$.

In the clip-level scenario, we no longer have access to the prior knowledge regarding sound class activities at a frame-level granularity, but we can similarly use clip-level weights if the sound classes are not uniformly distributed at the clip level. We did not consider this in our experiments as we assumed they were uniformly distributed at the clip level (equal probability of being active within a clip).

\subsection{Network Architecture}
\label{sec:net-arch}

\begin{figure}[tbp]
  \centering
 \includegraphics[width=8.8cm,height=6.8cm]{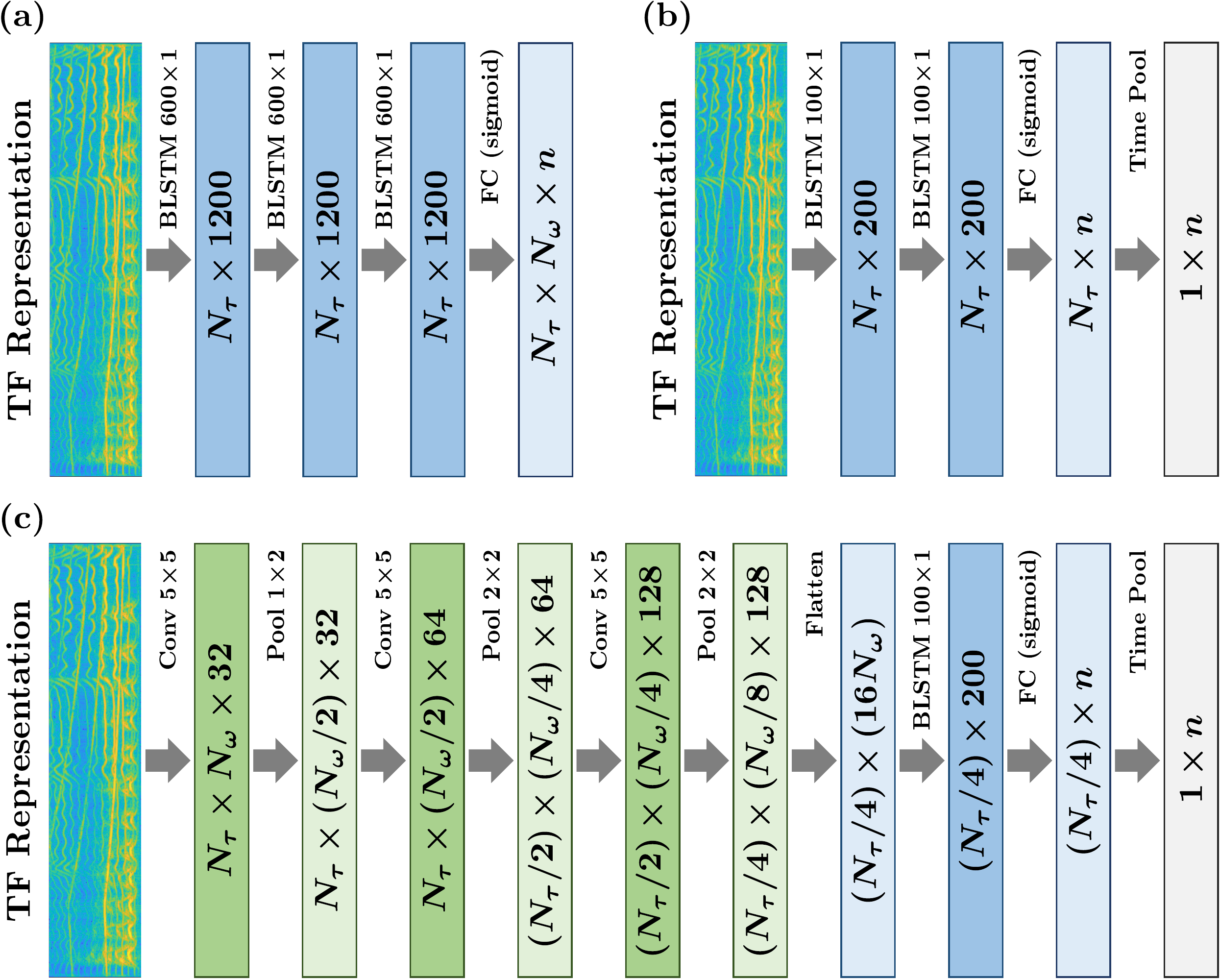}
\caption{Architectures of (a) the
separator, (b) the RNN classifier, and (c) the 2D-CRNN classifier. $N_{\tau}$ and $N_{\omega}$ denote the number of time frames and frequency bins in the input representation, respectively. $n$ is the total number of sound classes. }\vspace{-.4cm}
\label{fig:net_archs}
\end{figure}

The architecture of the separator block used in our experiments is depicted in Fig.~\ref{fig:net_archs}(a). It is composed of a 3-layer bidirectional long short-term memory (BLSTM) network, with each layer including 600 nodes in each direction. A fully connected layer maps the output of the BLSTM network to $n$ masks with the same size as the input mixture. Activation functions of all BLSTM units are \emph{tanh}, while the dense layer outputs go through \emph{sigmoid} functions, so that the mask values are always in [0,1].

To design a frame-level classifier, we explored a number of architectures, ranging from very simple, such as a small stack of fully connected layers, to increasingly more sophisticated ones, such as convolutional recurrent neural networks (CRNNs) \cite{cakir2017convolutional,adavanne2017sound}.
The clip-level classifier in this work is a simple extension of the frame-level classifier. 
It is built by adding a max-pooling operator to the output of the frame-level classifier for each sound class, in order to perform frame-level to clip-level mapping of sound presence probabilities.  More specifically, denoting by $p_{i,\tau}(\bm{Y})$ the prediction of our frame-level classifier at frame $\tau$ for sound class $i$ with an input spectrogram $\bm{Y}$, our clip-level classifier prediction $p_i(\bm{Y})$ for class $i$ and input $\bm{Y}$ is obtained as:
\begin{equation}
    p_i(\bm{Y}) = \max_{\tau}  p_{i,\tau}(\bm{Y}).
\end{equation}
We leave the investigation of separation performance for some of the more advanced temporal pooling operations explored in~\cite{mcfee2018adaptive} and \cite{wang2019comparison} to future work.

Here, we present the two architectures that performed best in our experiments: 
\begin{enumerate}[label=\roman*)]
    \item 
    \emph{RNN}:
    A 2-layer BLSTM network, with each layer including 100 nodes in each direction, followed by a fully connected layer that maps the BLSTM output for every time frame to $n$ class probabilities.
    Activation functions of all BLSTM units are again \emph{tanh}. Since the classifier is expected to detect the presence of multiple overlapping sound classes independently from one another, its output for each class is mapped to probability values through a \emph{sigmoid} function. Figure \ref{fig:net_archs}(b) illustrates this architecture.
    
    \item
    \emph{2D-CRNN}:
    A CRNN architecture composed of a 3-layer 2D convolutional network including max-pooling after each layer, followed by a BLSTM layer and a fully connected layer, which maps the BLSTM output to class probabilities. Activation functions of convolutional, BLSTM, and fully connected layers are \emph{relu}, \emph{tanh}, and \emph{sigmoid}, respectively. The output of each convolutional layer is batch normalized prior to the application of the activation function.
    Figure \ref{fig:net_archs}(c) illustrates this architecture in detail. This network is a slightly modified version of the SED model proposed in \cite{wang2019comparison}. 
    Note that the second and third pooling operations in the convolutional network are applied across both frequency and time axes, which results in a downsampled version of frame-level predicted probabilities. To match this coarser time resolution while computing the frame-level loss values, we also downsample the true weak labels via max-pooling.
\end{enumerate}

\section{Experiments}
\label{sec:experiments}

In this section, we present the results of our experiments, and compare our proposed weakly supervised method to the fully supervised approach using strong labels. We also discuss our observations regarding the importance of different model components and parameter setups.

\subsection{Sound Event Dataset}
\label{sec:dataset}

\emph{UrbanSound8K}\footnote{\href{https://urbansounddataset.weebly.com/urbansound8k.html}{https://urbansounddataset.weebly.com/urbansound8k.html}} \cite{SalamonUrbanSoundACMMM14} is a dataset of 8732 sound excerpts of length $\leq$ 4 s, taken from field recordings. The dataset contains 10 sound classes: air conditioner, car horn, children playing, dog bark, drilling, engine idling, gun shot, jackhammer, siren, and street music. The audio excerpts are labeled based on the sound classes to which they belong as well as their salience in the auditory scene (foreground or background).

In our experiments, five classes are included in mixture generation: car horn, dog bark, gun shot, jackhammer, and siren. The class selection was made
based on two criteria: i) 
audio examples in one class should contain mostly the sound corresponding to the class label, with a reasonable salience level, and ii)
audio examples from different classes should be acoustically distinct enough so that they are at least recognizable as different sounds by human listeners.
The air conditioner, children playing, and street music classes do not meet the first criterion, as their examples
contain target sounds that are either in the background and barely audible, or accompanied by sounds from other classes. The drilling, jackhammer, and engine idling classes include many examples that sound very similar, thus only one of them was selected.   

Audio mixtures in our dataset are 4 seconds long and sampled at 16 kHz. 
Each mixture is composed of at least one \emph{sound event} (i.e., a single occurrence of a sound class) from one of the five selected classes. 
The total number of sound events per mixture is sampled from a zero-truncated Poisson distribution with an expected value of $\lambda$. It is important to note that this number can include multiple sound events from one class, which are grouped together and regarded as one source while generating the weak labels. Thus, the value of $\lambda$ determines how crowded the auditory scene is. For instance, $\lambda=10$ means there are on average 10 sound events (from any class) per mixture. For each event, we first select one of the five classes uniformly at random, and then sample the actual sound event from all sounds of that class uniformly as well. Sound events are of arbitrary lengths, ranging from 0.5 s to 4 s, with a start time sampled uniformly at random under the constraint that the event fits entirely in the 4 s clip. The level of each sound event is randomly sampled from a uniform distribution of -30 to -25 loudness units full-scale (LUFS)~\cite{grimm2010toward}. 

\emph{UrbanSound8K} is distributed with the data split into 10 folds. We use folds 1-6 for creating the training set, folds 7-8 for the validation set, and folds 9-10 for the test set.  Our training, validation, and test sets include 20,000, 5000, and 5000 mixtures, respectively. The frame-level prior probabilities of activity $\gamma_i$ (see Section~\ref{sec:class_weights}) for the five sound classes and $\lambda$ values of 5 and 10 are presented in Table \ref{class-activation}.  Since all classes were sampled uniformly during training, the clip-level prior probabilities of activity are uniformly distributed and thus not reported.
To gain an idea of the amount of overlap between sources, we have also computed the distribution of frames and clips containing different numbers of sources in the entire training set. This information is provided in Table \ref{src-overlap}. 

\renewcommand{\arraystretch}{1}
\begin{table}[htbp]
\caption{Frame-level prior probabilities of activity $\gamma_i$ for the five selected sound classes. The probabilities are computed for training datasets with different $\lambda$ values. }\vspace{-.6cm}
\label{class-activation}
\begin{center}
\resizebox{\columnwidth}{!}{
\begin{tabular}{cccccc}
\toprule
 & \multicolumn{5}{c}{Sound class} \\
\cmidrule(lr){2-6}
$\lambda$ &  Car horn & Dog bark &Gun shot & Jackhammer & Siren\\
\midrule
5 &0.26&0.36&0.27&0.40 &0.40 \\
10&0.44&0.57&0.45&0.62&0.63\\
\bottomrule
\end{tabular}
}\vspace{-0.5cm}
\end{center}
\end{table}

\renewcommand{\arraystretch}{1}
\begin{table}[htbp]
\caption{Distribution of frames and clips containing different numbers of sources in training datasets with different $\lambda$ values. }\vspace{-.3cm}
\label{src-overlap}
\begin{center}
\begin{tabular}{ccccccc}
\toprule
 & \multicolumn{6}{c}{Number of sources per frame} \\
\cmidrule(lr){2-7}
$\lambda$ &  0 & 1 &2 & 3 & 4& 5\\
\midrule
5 & 0.17 &0.28&0.30&0.18&0.06&0.01 \\
10&0.07&0.13&0.21&0.28&0.23&0.08\\
\bottomrule
\toprule
& \multicolumn{6}{c}{Number of sources per clip} \\
\cmidrule(lr){2-7}
$\lambda$ &  0 & 1 &2 & 3 & 4& 5\\
\midrule
5 & 0.00 &0.06&0.20&0.34&0.30&0.10 \\
10&0.00&0.00&0.02&0.12&0.38&0.48\\
\bottomrule
\end{tabular}\vspace{-0.5cm}
\end{center}
\end{table}

\subsection{Training Setup}
\label{sec:features}
In all training sessions, we used the ADAM optimizer, with a learning rate of $10^{-4}$, $\beta_1 = 0.9$, and $\beta_2 = 0.999$. The batch size was set to 10 in all experiments, except in the experiment investigating the effect of a shorter window size (8 ms) where the batch size was set to 8 (see Table \ref{results-winlen}).  We train all networks until the loss on the validation set stops improving for five consecutive epochs, with a maximum of 50 epochs. The separator takes the log-magnitude STFT of a mixture as input using the square root of a Hann window of size 32 ms and a hop size of 8 ms. 
To provide an upper bound for the separation performance, we trained a separator network as described in Section \ref{sec:net-arch} on strong labels (i.e., target sources) with the weighted version of the fully supervised mask inference loss function $\mathcal{L}_{\mathrm{mi},W}$ in \eqref{loss-strong-src-weighted}.
In the weak label cases, we considered three training strategies: i) training the separator and classifier jointly from scratch using \eqref{frame_loss}, ii) pre-training the classifier until convergence using \eqref{frame_class_mix}, then training the separator through the pre-trained classifier while the classifier is being fine-tuned using \eqref{frame_loss}, and iii) pre-training the classifier until convergence using \eqref{frame_class_mix}, then training the separator through the pre-trained and fixed classifier using \eqref{frame_loss}, where only the term involving the estimated sources contributes to the gradient. Our most effective setup used the 2D-CRNN classifier shown in Fig.~\ref{fig:net_archs}(c), with linear magnitude STFT features as classifier input, where we first pre-trained the classifier on the mixtures, then trained the separator through the fixed pre-trained classifier using a mixture loss weight $\alpha=100$ in \eqref{frame_loss} and \eqref{clip_loss}. We use this as our default setup, and explore the importance of these choices in Section~\ref{sec:ablation}. 
 
\subsection{Sound Event Separation Results}
\label{sec:results}

We evaluate the performance of the classifier in terms of F-measure $\mathcal{F}=\frac{2\mathcal{P}\mathcal{R}}{\mathcal{P}+\mathcal{R}}$, the harmonic mean of precision $\mathcal{P}=\frac{\text{TP}}{\text{TP}+\text{FP}}$ and recall $\mathcal{R}=\frac{\text{TP}}{\text{TP}+\text{FN}}$, where $\text{TP}$, $\text{FP}$, and $\text{FN}$ respectively denote the number of true positives, false positives, and false negatives in the classification results. To measure the quality of the separated sources, we use the scale-invariant source-to-distortion ratio (SI-SDR) \cite{le2019sdr,Isik2016}, which has been shown to be more appropriate for single-channel instantaneous separation evaluation than the original SDR \cite{vincent2006performance}.  
When computing SI-SDR over the test set, we ignore silent sources as well as any mixtures that contain isolated sources, which can happen occasionally for $\lambda=5$ (see Table~\ref{src-overlap}).

Tables \ref{fmeasure-frame} and \ref{fmeasure-clip} present the average F-measure for frame-level and clip-level sound classification, respectively. The input to the RNN classifier is a magnitude mel spectrogram with 40 filters and the 2D-CRNN input is a magnitude STFT with linear frequency. It can be observed that, at the frame level, the 2D-CRNN classifier outperforms the RNN classifier by a large margin in identifying all sound sources. The two classifiers perform more similarly at the clip level, with 2D-CRNN working slightly worse than RNN for the jackhammer class, but still better than RNN for all other classes.

\renewcommand{\arraystretch}{1}
\begin{table}[tbp]
\caption{Frame-level sound source classification performance in terms of F-measure. The classifiers are trained and tested on datasets with $\lambda=5$. }\vspace{-.5cm}
\label{fmeasure-frame}
\begin{center}
\resizebox{\columnwidth}{!}{\setlength{\tabcolsep}{3pt}
\begin{tabular}{cccccc}
\toprule
  & \multicolumn{5}{c}{Sound class}  \\
\cmidrule(lr){2-6}
Classifier &  Car horn & Dog bark &Gun shot & Jackhammer & Siren \\
\midrule
RNN (mel-40) & 0.850 &  0.813 &  0.809 &  0.915 &  0.811 \\
2D-CRNN (STFT) & 0.948 &   0.870 &  0.856 &  0.940  & 0.876\\
\bottomrule
\end{tabular}
}\vspace{-0.3cm}
\end{center}
\end{table}

\renewcommand{\arraystretch}{1}
\begin{table}[tbp]
\caption{Clip-level sound source classification performance in terms of F-measure. The classifiers are trained and tested on datasets with $\lambda=5$.}\vspace{-.5cm}
\label{fmeasure-clip}
\begin{center}
\resizebox{\columnwidth}{!}{\setlength{\tabcolsep}{3pt}
\begin{tabular}{cccccc}
\toprule
  & \multicolumn{5}{c}{Sound class}  \\
\cmidrule(lr){2-6}
Classifier &  Car horn & Dog bark &Gun shot & Jackhammer & Siren \\
\midrule
RNN (mel-40) & 0.914   & 0.915  & 0.915   & 0.934 & 0.864 \\
2D-CRNN (STFT) & 0.958   &0.924 &  0.949 &  0.922 &  0.914\\
\bottomrule
\end{tabular}
}\vspace{-0.1cm}
\end{center}
\end{table}

\renewcommand{\arraystretch}{1}
\begin{table*}[htbp]
\caption{ Mean/median SI-SDR values (\db) for all sound classes and separators trained using different labels. $\Delta$SI-SDR indicates the SI-SDR improvement. The last column shows the results over all samples and all classes. The 2D-CRNN classifier is used in both weak label cases.
Models are trained and tested on datasets with $\lambda=5$. }\vspace{-.3cm}
\label{sep_stats}
\begin{center}
\resizebox{0.8\textwidth}{!}{
\begin{tabular}{l
S[table-format=2.2]@{\,\( / \)\,}S[table-format=3.1]
S[table-format=2.2]@{\,\( / \)\,}S[table-format=3.1]
S[table-format=2.2]@{\,\( / \)\,}S[table-format=3.1]
S[table-format=2.2]@{\,\( / \)}S[table-format=3.1]  
S[table-format=2.2]@{\,\( / \)\,}S[table-format=3.1]
S[table-format=2.2]@{\,\( / \)\,}S[table-format=3.1]}
\toprule
  & \multicolumn{10}{c}{Sound class} & \multicolumn{1}{c}{}  &\\
\cmidrule(lr){2-11}
 &\multicolumn{2}{c}{Car horn}& \multicolumn{2}{c}{Dog bark}& \multicolumn{2}{c}{Gun shot}& \multicolumn{2}{c}{Jackhammer}& \multicolumn{2}{c}{Siren} & \multicolumn{2}{c}{Overall}\\
\midrule
Input SI-SDR & -5.8 & -5.7 &  -5.4 & -5.4 & -5.5 & -5.8  & -2.9 &-2.8  & -3.0 & -3.0 & -4.5 & -4.5 \\
$\Delta$SI-SDR-strong & 9.9 & 7.9 & 10.0 & 9.2   &  12.5 & 11.2  & 7.8 & 6.8 & 4.9 & 6.2 & 9.0 & 8.3 \\
$\Delta$SI-SDR-frame & 7.0 & 5.4 & 8.3 & 7.7 & 9.7 & 9.1 & 5.7 & 4.9 & 3.1 & 4.3 & 6.8 & 6.2\\
$\Delta$SI-SDR-clip & 6.5 & 5.7 & 6.4 & 6.1 & 8.8 & 8.3 & 4.6 & 4.0 & 1.8 & 3.5 & 5.6 & 5.5\\
\bottomrule
\end{tabular}
}\vspace{-0.3cm}
\end{center}
\end{table*}

\begin{figure*}[htbp]
  \centering
 \includegraphics[width=0.9\textwidth]{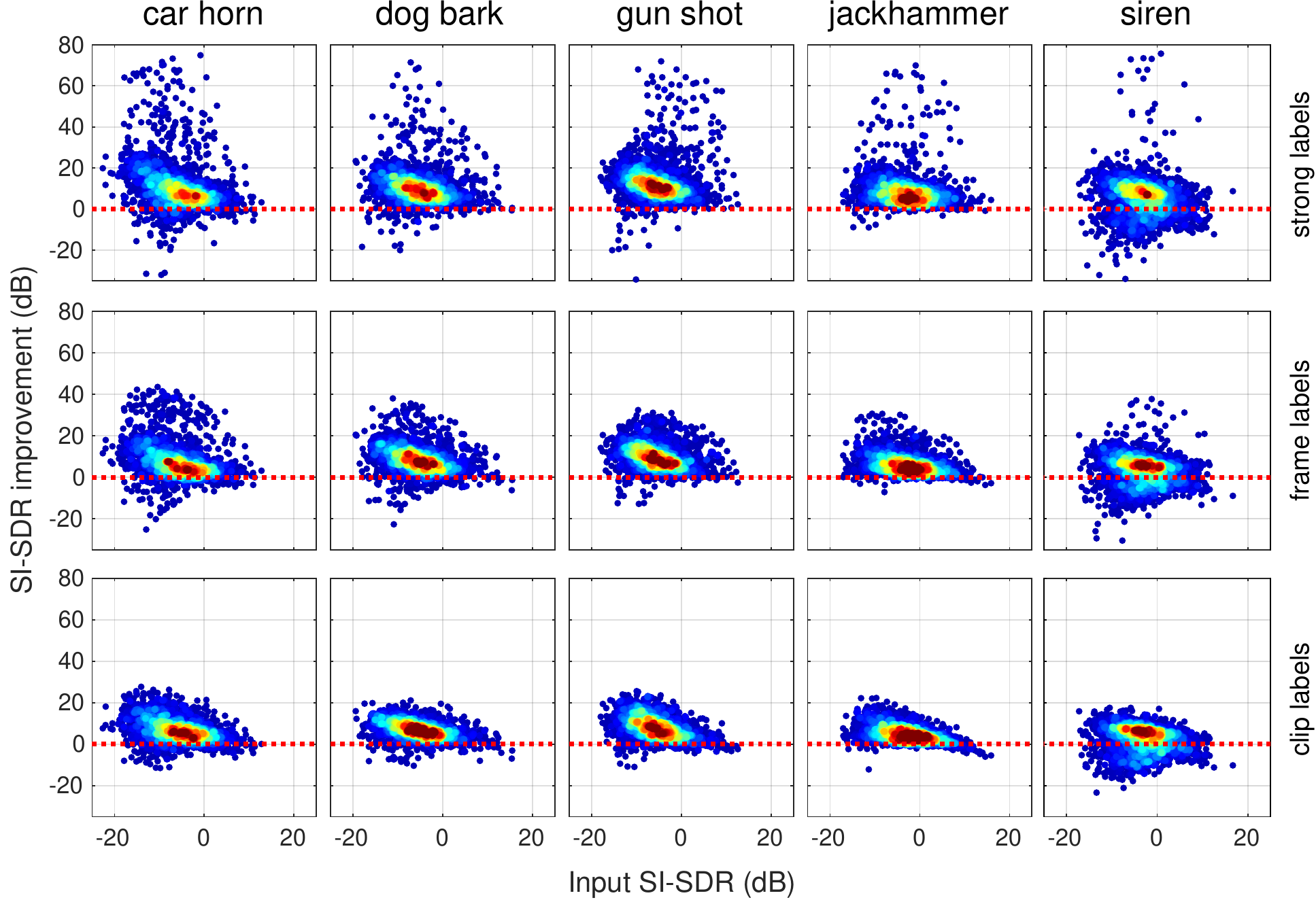}\vspace{-0.3cm}
\caption{Separation results for all sound classes when the 
separator is trained on strong  labels (top row), frame-level weak labels (middle row), and clip-level weak labels (bottom row). All panels show SI-SDR improvement versus input SI-SDR values. The 2D-CRNN classifier and the magnitude STFT input are used in experiments with both frame-level and clip-level labels. Warmer colors mean higher densities of data points. }
\label{fig:sdr-imp-in}\vspace{-0.3cm}
\end{figure*}

Source separation results for strong labels (fully supervised upper bound) and weak labels are shown in Table \ref{sep_stats} and Fig.~\ref{fig:sdr-imp-in}, where the weak label results are obtained with our default setup described above, using a pre-trained 2D-CRNN classifier.  
In Table \ref{sep_stats}, we present both means and medians of SI-SDR values, since the former measure is commonly used for reporting separation results in the literature, and the latter is better suited to the non-Gaussian distributions with a large number of outliers, which is particularly the case for the strong label results in Fig.~\ref{fig:sdr-imp-in}. 
Since the overall trends between mean and median results in Table~\ref{sep_stats} are similar, for clarity we report only mean results for the ablation studies in Section~\ref{sec:ablation}.
From these summary statistics, we see SI-SDR improvements with respect to the mixture for all classes with both frame- and clip-level weak labels.  The smallest and largest SI-SDR improvements in Table \ref{sep_stats} are for the siren and gun shot classes, respectively.  The siren class in our dataset contains a more diverse set of sounds compared to other classes (e.g., police siren versus ambulance siren), which is likely the reason why it is the most difficult sound type to separate even when strong labels are used.

The scatter plots of separation results, shown in Fig.~\ref{fig:sdr-imp-in}, allow a more detailed comparison between the performance of separators trained through different types of labels. We note that all test mixtures included in these plots contain at least two sound sources.
Each panel shows the amount of SI-SDR improvement versus input SI-SDR for all test set examples of one sound class. The input SI-SDR refers to the SI-SDR obtained when considering the input mixture as the estimate for the target source.  
One common trend observed in all cases is the downward tilted shape of the data distribution, which is also typically observed in speech separation \cite{Isik2016,Wichern2019Interspeech09}. This pattern  indicates that the highest SI-SDR improvement is achieved for low-SI-SDR inputs and the amount of improvement shrinks when using inputs with higher SI-SDR values. 

When going from strong to weak labels in all sound classes, an obvious trend is a decrease in the number of points in the higher end of SI-SDR improvement values. 
For example, in the plot corresponding to the results for the car horn class and strong labels (top row, leftmost panel), there are several points with SI-SDR improvements above 50 dB. When using frame-level labels, the highest SI-SDR improvement drops to around 40 dB, and it decreases even further down to 30 dB when using clip-level labels. Interestingly, however, the high-density regions of the distributions in each class seem to remain largely similar, contrary to what one may have expected given the difference in the strength of labels used for training.  
Although frame-level labels yield better results than clip-level labels in general, the distribution of output SI-SDRs for these two label types seems to be very similar 
in all cases. Furthermore, both weak-label distributions seem to have large
amounts of overlap with strong-label distributions and to provide significant SI-SDR improvement over the input SI-SDRs.

\subsection{Ablation studies}\label{sec:ablation}
\subsubsection{Training strategies}
The effect of different training strategies
on separation performance can be observed in Table~\ref{training-strategy}. The joint separation-classification model was trained on frame-level weak labels under the three training strategies listed in Section \ref{sec:jsc-model}.  Regardless of the classifier architecture, the best separation results are achieved when the classifier is pre-trained on mixtures and its parameters are then fixed when training the separator. Training the separator and classifier jointly from scratch, or fine-tuning the classifier when the separator is being trained always resulted in a worse separation performance in our experiments. We hypothesize that this behavior is due to the co-adaptation of the two networks, where the classifier can adapt its weights to correctly classify errors made by the separator, rather than forcing the separator to output estimated sources that match the previously learned representation for each sound class.  In other words, this co-adaptation weakens the ability of the classifier to objectively assess the performance of the separator. 

\begin{table}[htbp]\vspace{-.1cm}
\caption{Mean SI-SDR improvement (\db) for different training strategies, over all classes. The models are trained on frame-level labels. In all cases, $\alpha = 100$, $\lambda = 5$, and the average input SI-SDR is $-4.5$ \db.} \vspace{-.4cm}
\label{training-strategy}
\begin{center}
\resizebox{0.95\columnwidth}{!}{
\begin{tabular}{cccc}
\toprule
&\multicolumn{3}{c}{Training strategy} \\
\cmidrule(lr){2-4}
Classifier & Joint & Fine-tune classifier & Fix classifier\\
\midrule
RNN (mel-40) & $-4.4$  & 5.5 & \textbf{6.2} \\
2D-CRNN (STFT) &$-0.2$ & 1.3  & \textbf{6.8}  \\
\bottomrule 
\end{tabular}
}\vspace{-.1cm}
\end{center}
\end{table}

\subsubsection{Mixture loss}
To investigate the effect of the mixture loss term, we trained the separator network using different $\alpha$ values in the overall frame-level loss of \eqref{frame_loss}. The SI-SDR improvement results, presented in Table \ref{results-alpha}, 
clearly show the importance of this loss term for the separation task. A similar trend is observed for both classifiers. When $\alpha=0$, only the classification loss is used to train the separator, which leads to poor separation performance as the separator network only needs to isolate the TF features necessary for classification, not signal reconstruction.  Conversely,  a comparatively very low contribution of the classification loss term (e.g., $\alpha = 10^4$) results in degraded performance as the separator only needs to reconstruct the mixture without isolating the individual sound sources.  A good balance between the two loss terms (e.g., $\alpha = 10^2$), is essential to obtain high SI-SDR gains.

\begin{table}[htbp]\vspace{-.1cm}
\caption{Mean SI-SDR improvement (\db) using different mixture loss weights, over all classes. The models are trained on frame-level labels. In all cases, $\lambda = 5$ and the average input SI-SDR is $-4.5$ \db.  } \vspace{-.4cm}
\label{results-alpha}
\begin{center}
\resizebox{.9\columnwidth}{!}{
\begin{tabular}{cccccc}
\toprule
& \multicolumn{5}{c}{Mixture loss weight ($\alpha$)} \\
\cmidrule(lr){2-6}
Classifier & $0$ & $10$ & $10^2$ & $10^3$ & $10^4$\\
\midrule
RNN (mel-40) & 1.9  & 4.6 & \textbf{6.2} & 5.3  & 1.9\\
2D-CRNN (STFT) & 0.9 & 3.9  & \textbf{6.8} & 5.1& 1.1 \\
\bottomrule
\end{tabular}
}\vspace{-.1cm}
\end{center}
\end{table}

\subsubsection{TF representation}
The properties of the audio representation input to the classifier, such as frequency scaling and resolution, proved to have a considerable impact on the separation results in our experiments. 
The performance of the separator is essentially correlated with the efficacy of the classifier in capturing the spectro-temporal patterns that distinguish each sound class from the others. For instance, a classifier that depends only on a few frequency bins to identify a sound will output accurate class probabilities as long as the separator assigns correct amounts of energy to those bins. Using such a classifier, the model could correctly identify an impulsive, broadband sound (e.g., gun shot) in a mixture, even if the separated source estimate includes only a small portion of the actual spectral content.

One way to address this problem is to force the classifier to produce predictions based on broader frequency ranges by decreasing the frequency resolution of mixtures and estimated sources prior to feeding them to the classifier. 
To lower the frequency resolution, we consider applying a mel-frequency filterbank to the magnitude STFTs to be used as classifier inputs. The mel-frequency filterbank also has the advantage of changing the frequency resolution logarithmically, with a grid that is finer across lower frequencies, maintaining most of the information necessary to distinguish harmonic sources, and grows coarser as the frequency increases. We investigate the effect of frequency scaling and  resolution on the quality of spectral patterns learned by the classifier, which in turn impacts the separation quality, by using two different representations as the classifier input: a linear magnitude STFT and a linear magnitude mel spectrogram with a varying number of mel filters. The STFT parameters (window size and hop length) are the same for the separator and classifier inputs. 
 
The amount of SI-SDR improvement for different mel-frequency filterbanks (featuring different numbers of filters and different center frequencies) are provided in Table \ref{freq-scale}. The results for the linear magnitude STFT, corresponding to the linear frequency case (no filterbank), are reported as ``Linear Freq.'' in the table.
As can be seen, changing the frequency scale and resolution of the classifier input can make a difference of up to 2 dB in the average SI-SDR improvement. The performance of a model using the RNN classifier can be improved up to 0.4 dB by using a mel spectrogram as input. The best number of mel filters, however, seems to be difficult to choose without running a grid search. The 2D-CRNN classifier, on the other hand, provides the highest improvement when the original magnitude STFT is used as input. 
We hypothesize that since convolutional networks inherently downsample frequency, using an input with low frequency resolution (e.g., mel) is more harmful than beneficial, while the RNN architecture, which performs no implicit downsampling, benefits from using mel spectrogram inputs.  

We also note that the 2D-CRNN classifier consistently outperforms the RNN classifier in terms of separation performance in Tables~\ref{training-strategy}-\ref{freq-scale}.  Since the 2D-CRNN classifier also provided the best classification performance for most classes in Tables~\ref{fmeasure-frame} and~\ref{fmeasure-clip}, these results imply that better classification performance is correlated with better separation performance when training a weakly labeled separation system.  Further refinements of the classifier may thus lead to improved separation quality.

\begin{table}[htbp]
\caption{Mean  SI-SDR improvement (\db) using different frequency scales and resolutions, over all classes. The models are trained on frame-level labels. In all cases, $\alpha=100$, $\lambda = 5$ and the average input SI-SDR is $-4.5$ \db.  } \vspace{-.2cm}
\label{freq-scale}
\begin{center}
\resizebox{.93\columnwidth}{!}{
\begin{tabular}{ccccccc}
\toprule
& \multicolumn{5}{c}{Number of mel filters} & \\
\cmidrule(lr){2-6}
Classifier & 10 & 20 & 30 & 40 & 56 & Linear freq.\\
\midrule
RNN  & 5.5  & \textbf{6.4} & 6.0 & 6.2 & 5.0 & 6.1\\
2D-CRNN  & 5.3 & 4.8 & 5.9 & 6.0 & 6.2  & \textbf{6.8} \\
\bottomrule
\end{tabular}
}\vspace{-0.3cm}
\end{center}
\end{table}

\begin{table}[htbp]
\caption{Mean SI-SDR improvement (\db) for different STFT window sizes, over all classes. In all cases, the 2D-CRNN classifier is used with magnitude STFT input features. The models are trained on frame-level labels. In all cases, $\alpha = 100$, $\lambda = 5$, the overlap between windows is 75\%, and 
the average input SI-SDR is $-4.5$ \db. } \vspace{-.4cm}
\label{results-winlen}
\begin{center}
\resizebox{\columnwidth}{!}{\setlength{\tabcolsep}{3pt}
\begin{tabular}{cccccc}
\toprule
  & \multicolumn{5}{c}{Sound class}  \\
\cmidrule(lr){2-6}
Win. size &  {Car horn} & {Dog bark} &{Gun shot} & {Jackhammer} & {Siren} \\
\midrule
\phantom{1}8 ms &5.8 &6.8&10.0&4.9&3.0 \\
16 ms &\textbf{7.8}&\textbf{8.7}&\textbf{10.4}&\textbf{6.1}&\textbf{4.1}\\
32 ms&7.0&8.3&\phantom{1}9.7&5.7&3.2\\
64 ms&5.1&5.8&\phantom{1}7.0&4.1&2.8\\
\bottomrule
\end{tabular}
}
\end{center}
\end{table}

We further investigate the effect of frequency and time resolutions by varying the window size of the magnitude STFTs input to both separator and classifier blocks. We run this experiment using the best performing model thus far, which includes a 2D-CRNN classifier with magnitude STFT (linear frequency scale) as input. 
The results, provided in Table \ref{results-winlen}, demonstrate that decreasing the time resolution by using windows longer than 32 ms degrades the results, while decreasing the window size provides performance improvement.
The limit on the improvement, however, seems to be reached when the window size is 16 ms, as using a window size of 8 ms results in worse performance, likely due to the poor frequency resolution.  

\begin{table*}[htbp]
\caption{Mean SI-SDR improvement (\db) for separators trained using strong labels, frame-level weak labels, clip-level weak labels, datasets with different $\lambda$ values, and weighted (left four columns) or not weighted (right four columns) loss functions. All separators are trained using the 2D-CRNN classifier. } \vspace{-.2cm}
\label{sdr-lambda}
  \sisetup{table-format=2.1,round-mode=places,round-precision=1,table-number-alignment = center,detect-weight=true,detect-inline-weight=math}
\begin{center}
\resizebox{.7\textwidth}{!}{
\begin{tabular}{lSSSSSSSS}
\toprule
&\multicolumn{4}{c}{With class weights} & \multicolumn{4}{c}{Without class weights}\\
\cmidrule(lr){2-5} \cmidrule(lr){6-9}
Training $\lambda$  & \multicolumn{2}{c}{5} & \multicolumn{2}{c}{10}&\multicolumn{2}{c}{5}&\multicolumn{2}{c}{10} \\
\cmidrule(lr){2-3}\cmidrule(lr){4-5} \cmidrule(lr){6-7}\cmidrule(lr){8-9}
Testing $\lambda$ & 5 & 10 & 5 & 10 &5 & 10 & 5 & 10\\
\midrule
Input SI-SDR & -4.5  & -6.2  & -4.5 &-6.2 & -4.5  & -6.2  & -4.5 &-6.2 \\
$\Delta$SI-SDR-strong & 9.0  & 7.0 & 9.4 & 7.3& 3.2  & 1.8 & 9.3 & 7.4\\
$\Delta$SI-SDR-frame & 6.8  & 5.2 & 6.4 & 5.4 & 6.3  & 4.9 & 6.1 & 5.3\\
$\Delta$SI-SDR-clip & 5.6  & 4.7 & 4.9 & 4.4 & 5.6  & 4.7 & 4.9 & 4.4\\
\bottomrule
\end{tabular}
}\vspace{-.4cm}
\end{center}
\end{table*}

\subsubsection{Source density}
Next, we compare how the density of sound sources in the scene impacts network performance.  As mentioned in Section~\ref{sec:dataset}, the parameter $\lambda$ used when creating our mixtures determines the expected number of events in each four-second scene.  The left four columns in Table~\ref{sdr-lambda} compare separation performance between training using strong labels (fully supervised upper bound) and both frame- and clip-level weak labels for different $\lambda$ values, where we only report results with the 2D-CRNN for brevity.  In the fully supervised case, training with more difficult mixtures ($\lambda=10$) leads to improved separation performance compared to training with easier mixtures ($\lambda=5$).  However, for frame-level weak labels, training with $\lambda=10$ leads to slightly worse performance than training with $\lambda=5$, and for clip-level weak labels training with $\lambda=10$ causes a larger performance drop.  Revisiting Table~\ref{src-overlap}, we see that when $\lambda=10$ the training set contains no clips with only a single active source, compared to 6\% of the clips for $\lambda=5$, while there are some single source frames for both $\lambda$ values.  Therefore, we hypothesize that the drop in clip-level weak-label performance for the $\lambda=10$ training set in Table~\ref{sdr-lambda} is due to the lack of any training data containing labeled regions with isolated sources.  While the higher SI-SDR numbers in Table~\ref{sdr-lambda} for both frame-level and clip-level  weak labels using the $\lambda=5$ training set indicate that the network does likely make use of labeled regions containing isolated sources, the method can still work in their absence,
as shown by the SI-SDR improvements of 4.9 dB and 4.4 dB for the most difficult case of clip-level weak-labels and the $\lambda=10$ training set.

\subsubsection{Class weights}

Finally, we investigate the importance of the loss weights, computed based on the prior probability of sound class activities, in separation training (see Section \ref{sec:class_weights}). In our training setup, the loss weights are used in the 
case of strong labels and frame-level labels, in the formulation of the mask inference loss and the classification loss, respectively. 
The results presented in Table \ref{sdr-lambda} (right four columns) are obtained using the same setup as in the previous section, with the only difference that the loss terms are not weighted. Removing the weights when a less dense training dataset (e.g., $\lambda=5$) is used results in a larger performance drop in both strong and frame-level cases. Intuitively, this behavior is expected as in such scenarios, the prior probabilities are much smaller than 0.5 for sparser classes, such as gun shot (see Table \ref{class-activation}), and hence these classes are assigned larger weights than less sparse classes, such as jackhammer. However, as the prior probabilities get closer to 0.5 in a dataset with $\lambda=10$, the difference between class weights becomes smaller and their effect on separation results less noticeable. Further, it can be observed that removing the weights has a more dramatic impact on the strong label than frame-level label results. This can be explained by considering the fact that in the frame-level label case, the weights are only incorporated in the classification loss. 
Therefore, as long as the classification performance does not degrade significantly (which was the case in our experiments), the separation performance is anticipated not to vary dramatically.

\renewcommand{\arraystretch}{1}
\begin{table*}[htbp]
\caption{Mean/median SI-SDR values (\db) for MUSDB18 using strong and weak labels. $\Delta$SI-SDR indicates the SI-SDR improvement. The last column shows the results over all samples and all classes. }\vspace{-.3cm}
\label{musdb_table}
\begin{center}
\resizebox{0.7\textwidth}{!}{
\begin{tabular}{l
S[table-format=2.2]@{\,\( / \)\,}S[table-format=3.1]
S[table-format=2.2]@{\,\( / \)\,}S[table-format=3.1]
S[table-format=2.2]@{\,\( / \)\,}S[table-format=3.1]
S[table-format=2.2]@{\,\( / \)}S[table-format=3.1]  
S[table-format=2.2]@{\,\( / \)\,}S[table-format=3.1]
S[table-format=2.2]@{\,\( / \)\,}S[table-format=3.1]}
\toprule
  & \multicolumn{6}{c}{Sound class} & \multicolumn{1}{c}{}  &\\
\cmidrule(lr){2-7}
 &\multicolumn{2}{c}{Bass}& \multicolumn{2}{c}{Drums}& \multicolumn{2}{c}{Vocals}&  \multicolumn{2}{c}{Overall}\\
\midrule
Input SI-SDR & -4.3 & -3.9 &  -1.8 & -2.1 & -4.9 & -4.0  & -3.7 &-3.6  \\
$\Delta$SI-SDR-strong & 6.2 & 6.6 & 6.0 & 6.2   &  13.6 & 13.9  & 8.6 & 8.1 \\
$\Delta$SI-SDR-clip & 1.5 & 2.1 & 2.7 & 2.7 & 8.9 & 8.8 & 4.4 & 3.6 \\
\bottomrule
\end{tabular}
}\vspace{-0.3cm}
\end{center}
\end{table*}

\subsection{Unsuccessful attempts}
In addition to using the RNN and 2D-CRNN sound event classifiers shown in Figs.~\ref{fig:net_archs}(b) and~\ref{fig:net_archs}(c) for frequency pooling, we explored simple frequency pooling (e.g., average pooling over frequency), a learned linear transform, or a feedforward deep network (without memory).  In all cases, these frequency pooling approaches failed to learn to separate.  Furthermore, we experimented with an ``idempotent'' loss, where an additional loss function term enforced the separation network to pass estimated separated sources unchanged. We found that this loss term only hurt separation performance. Our hypothesis is that in most cases this constraint was redundant with information provided by the weak labels.  Finally, we explored using log magnitude STFT features (as opposed to linear magnitude) as input to the classifier.
Although log features gave slightly better classification performance, 
our separator networks did not train reliably, even when regularizing the 
log (i.e., adding a small positive value to the log input).

\subsection{Application to Music Source Separation}\label{sec:music_results}

\subsubsection{Dataset}  MUSDB18~\cite{musdb18} currently serves as the standard dataset for benchmarking music source separation algorithms~\cite{stoter19}.  It consists of 150 songs split into a 100 song training set and 50 song test set.  Each song consists of a mixture, and four stems that compose the mixture, containing isolated audio tracks for \textit{bass, drums, vocals} and \textit{other} source classes. These isolated sources serve as training targets (i.e., strong labels) for fully-supervised separation.  The lack of precision in the definition of the \textit{other} class makes training a classifier to reliably predict the presence or absence of this class difficult, so in the present study we remove this source and consider three-source mixtures of \textit{bass, drums}, and \textit{vocals}.

We further divide the training set into a subset of 86 songs used for training and another with the remaining 14 songs used as a validation set.  Each song in the training and validation sets is further split into four second non-overlapping chunks (matching the clip-length of our sound event dataset described in Section~\ref{sec:dataset}), which are used as input to our system.  The stems for each four-second chunk are used as strong labels, and the presence or absence of energy in that stem is used as a clip-level weak label. We only consider clip-level weak labels for MUSDB18 
since several of the stems in MUSDB18 contain bleed from other instruments or high levels of background noise that make obtaining accurate frame-level labels ambiguous. 

All audio is downsampled to 16 kHz, and our training setup matches exactly the best performing setup used for the sound event dataset described in Section~\ref{sec:features}, i.e., 2D-CRNN classifier with STFT features and mixture loss weight $\alpha=100$.  We made two modifications to the network architecture described in Section~\ref{sec:net-arch} for MUSDB: i) as mentioned in~\cite{mcfee2018adaptive} because music sources tend to be consistently active/inactive over periods of several seconds, we replaced the max-pooling operation at the output of our clip-level classifier with an average pooling operation, and ii) we add a batch normalization layer at the input of the separation network, which while not beneficial for urban sound event separation improved performance for MUSDB18.

\subsubsection{Results}  
\begin{figure}[htbp]
  \centering
 \includegraphics[width=1\columnwidth]{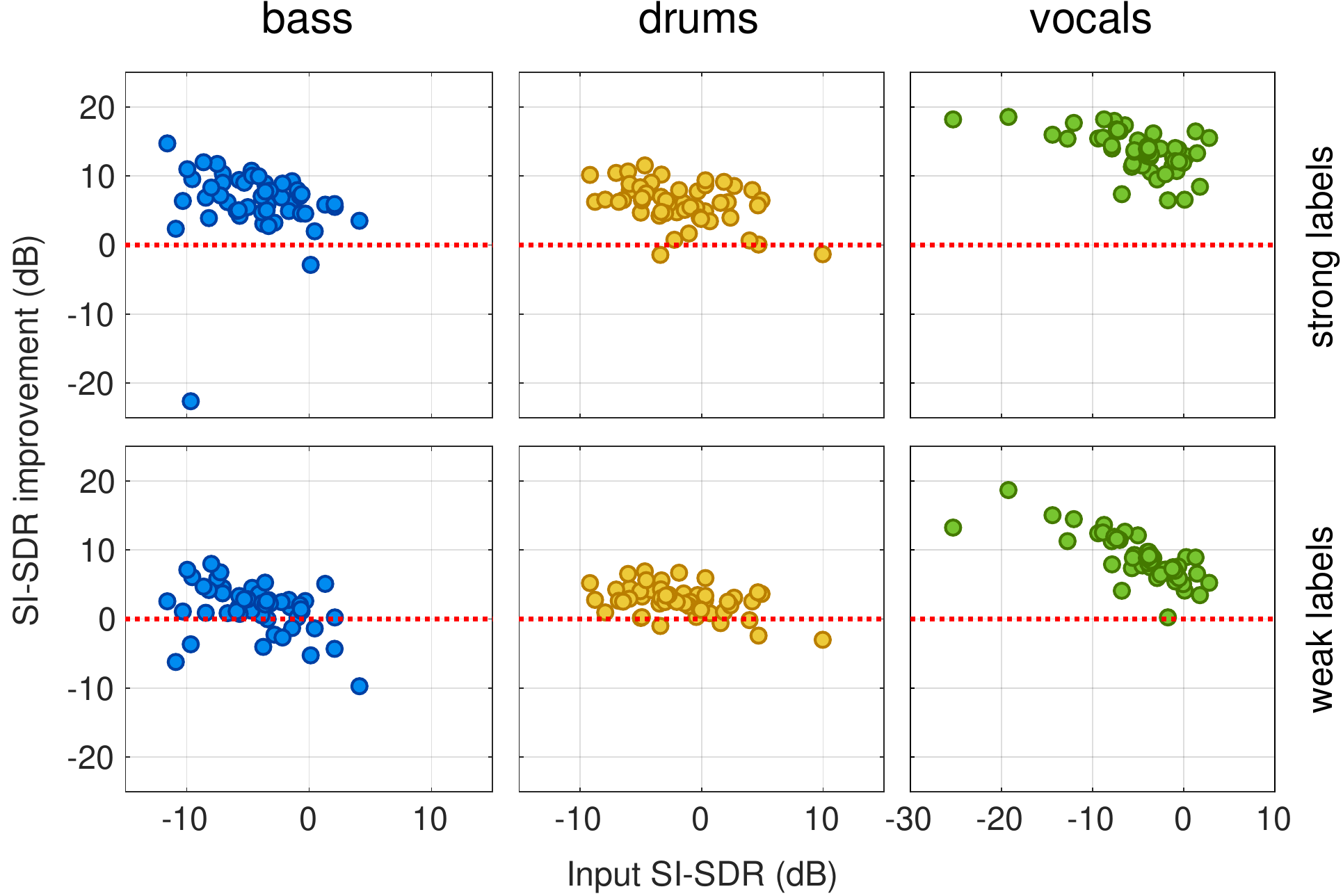}\vspace{-0.3cm}
\caption{Separation results for all sound classes in music mixtures when the 
separator is trained on strong  labels (top row) and clip-level labels (bottom row). All panels show SI-SDR improvement versus input SI-SDR values.
Each panel contains 50 data points, each point corresponding to an entire test song.
The 2D-CRNN classifier and the magnitude STFT input are used in experiments with clip-level labels.}
\label{fig:musdb_plot}\vspace{-0.3cm}
\end{figure}

We evaluate performance by computing SI-SDR on each of the 50 songs in the test set. 
We run each entire song through the separator network, i.e., we do not divide the songs in the test set into chunks.
Table~\ref{musdb_table} compares the overall performance in terms of mean and median SDR for strong labels and clip-level weak labels, while Fig.~\ref{fig:musdb_plot} shows corresponding scatter plots.  From Table~\ref{musdb_table}, 
we observe improvements in SI-SDR for all source classes and note that the 4.2 dB difference between the overall mean SDR improvements for the strong and clip-level weak labels (8.6 dB vs. 4.4 dB) for MUSDB18 is comparable to the 3.4 dB difference observed for sound events
in Table~\ref{sep_stats}.  We also note similarities in the overall shape of the strong and weak-label scatter plots of Fig.~\ref{fig:musdb_plot}, especially for vocals.  

Compared to vocals, both bass and drums exhibit a larger drop in performance between the strong and weak label cases.  This is most likely due to these sources almost always playing at the same time. Thus, without any prior information it is extremely difficult for a classifier trained on mixtures to learn the difference between these co-occurring classes. We also mention that when trying to include an overly broad class in our experiments, such as \textit{other} in MUSDB18, our separation networks tended to put most of the energy into the output corresponding to this class.  Tackling this issue and improving performance for consistently co-occurring sources such as bass and drums are important extensions of the current work.  However, given the difficult copyright issues necessary to obtain isolated sources for training music separation networks, we believe the results presented here represent a potentially important path in scaling up music source separation.
 
\section{Conclusion and future works}
\label{sec:conclusion}
We have presented an algorithm for training a source separation system with weak labels, where isolated sources are not required for the training process.  In our proposed model, an SED classifier is employed as the principal metric for loss calculation while training the separator. 
The model is trained to minimize an objective function that 
requires the separator to produce source estimates that are identifiable by the classifier.
The objective function also enforces the estimated sources to sum up to the mixture. Our experiments yielded promising results and showed significant SI-SDR improvement even when using weak labels on a very coarse-resolution time grid.

In the present work, we only explored a source separation algorithm based on magnitude masking in the spectral domain.  Moving forward, we could extend our weak label separation objectives to systems based on phase sensitive masking~\cite{Erdogan2015}, complex masking~\cite{williamson2016complex}, phase estimation~\cite{leroux2019phasebook}, time domain separation~\cite{luo2019convTasNet}, and/or deep TF embeddings~\cite{hershey2016deep}.  Combining the discriminative approach presented here with the generative approaches of \cite{stoller2018adversarial,michelashvili2019semi,karamatli2019audio} while still minimizing the amount of required supervision is also a potential avenue for future exploration. Finally, this work only considered mixtures of labelled sounds from a given set of classes, whereas real-world sound mixtures are likely to also contain unlabelled sounds from other classes. Moreover, we considered all instances of the same class as a single source, whereas one may in general be interested in further separating each instance. Dealing with such limitations is an important topic for future work. 

\vspace{-0.2cm}

\section*{Acknowledgment}

The authors would like to thank Shrikant Venkataramani, Dr.~Prem Seetharaman, and Ethan Manilow for helpful discussions and comments.

\balance

\bibliographystyle{IEEEtran}
\bibliography{refs}

\ifCLASSOPTIONcaptionsoff
  \newpage
\fi

\begin{IEEEbiography}
[{\includegraphics[width=1in,height=1.25in,clip,keepaspectratio]{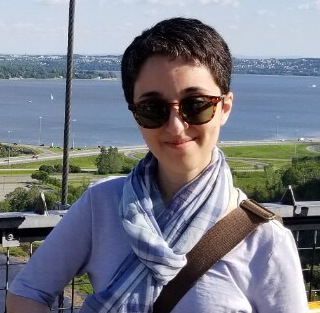}}]{Fatemeh Pishdadian}
is a Ph.D.\ candidate in Computer Science at Northwestern University.
She received her B.Sc.\ in Electrical Engineering from Ferdowsi University, Mashhad, Iran, and her M.Sc.\ in Electrical and Computer Engineering from George Mason University. Her research interest lies in the application of signal processing and machine learning methods to the analysis of audio/music. More specifically, the focus of her doctoral research has been on audio/music source separation. 
\end{IEEEbiography}

\begin{IEEEbiography}
[{\includegraphics[width=1in,height=1.25in,clip,keepaspectratio]{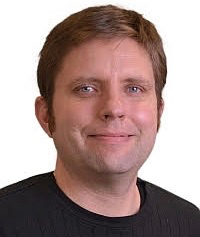}}]{Gordon Wichern}
is a Principal Research Scientist at Mitsubishi Electric Research Laboratories (MERL) in Cambridge, Massachusetts. He received his B.Sc.\ and M.Sc.\ degrees from Colorado State University in electrical engineering and his Ph.D.\ from Arizona State University in electrical engineering with a concentration in arts, media and engineering, where he was supported by a National Science Foundation (NSF) Integrative Graduate Education and Research Traineeship (IGERT) for his work on environmental sound recognition.  He was previously a member of the research team at iZotope, inc.\ where he focused on applying novel signal processing and machine learning techniques to music and post production software, and a member of the Technical Staff at MIT Lincoln Laboratory where he worked on radar signal processing.  His research interests include audio, music, and speech signal processing, machine learning, and psychoacoustics.
\end{IEEEbiography}

\begin{IEEEbiography}
[{\includegraphics[width=1in,height=1.25in,clip,keepaspectratio]{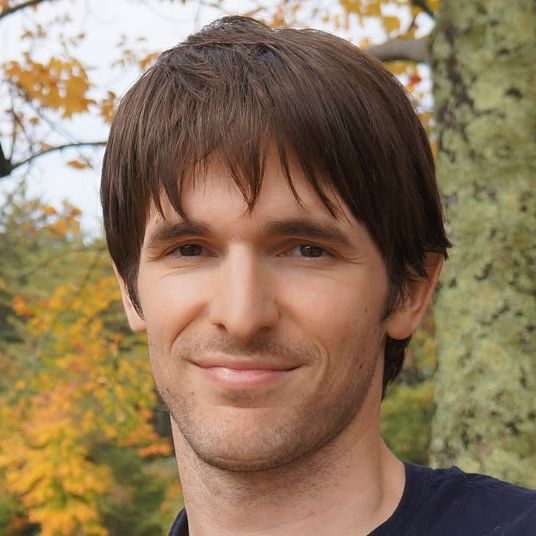}}]{ Jonathan Le Roux}
 is a Senior Principal Research Scientist and the Speech and Audio Senior Team Leader at Mitsubishi Electric Research Laboratories (MERL) in Cambridge, Massachusetts. He completed his B.Sc.\ and M.Sc.\ degrees in Mathematics at the Ecole Normale Sup\'erieure (Paris, France), his Ph.D.\ degree at the University of Tokyo (Japan) and the Universit\'e Pierre et Marie Curie (Paris, France), and worked as a postdoctoral researcher at NTT’s Communication Science Laboratories from 2009 to 2011. His research interests are in signal processing and machine learning applied to speech and audio. He has contributed to more than 100 peer-reviewed papers and 20 granted patents in these fields. He is a founder and chair of the Speech and Audio in the Northeast (SANE) series of workshops, and a Senior Member of the IEEE.
\end{IEEEbiography}

\end{document}